# Optical polarization based logic functions (XOR or XNOR) with nonlinear Gallium nitride nanoslab

F.A.Bovino[1,*], M.C.Larciprete[2], M.Giardina[1], A. Belardini[2], C. Sibilia[2], M. Bertolotti[2], A.Passaseo[3], V.Tasco[3].

[1]*Quantum Optics Lab. , Elsag-Datamat Via Puccini 2 Genova, Italy*
[2] *Dipartimento di Energetica, Università di Roma La sapienza, Via A.Scarpa 16 Roma, Italy*
[3]*CNR-NNL-INFM Unità di Lecce - Via per Arnesano Lecce, Italy*
*Corresponding author: fabio.bovino@elsagdatamat.com

**Abstract:** We present a scheme of XOR/XNOR logic gate, based on non phase-matched noncollinear second harmonic generation from a medium of suitable crystalline symmetry, Gallium nitride. The polarization of the noncollinear generated beam is a function of the polarization of both pump beams, thus we experimentally investigated all possible polarization combinations, evidencing that only some of them are allowed and that the nonlinear interaction of optical signals behaves as a polarization based XOR. The experimental results show the peculiarity of the nonlinear optical response associated with noncollinear excitation, and are explained using the expression for the effective second order optical nonlinearity in noncollinear scheme.

## 1. Introduction

In the near future all-optical signal processing will be necessary to build high-speed large-capacity optical communication networks and to avoid inefficient and cumbersome optical-electrical–optical (OEO) conversions. Among several fundamental optical logic gates which are basic functions in all-optical signal processing, XOR is fundamental for set of critical functions, such as label or packet switching, parity checking, decision making, address and head recognition, basic or complex computing, optical generation of pseudo-random patterns. The use of polarization states has been investigated since many years [1] and has been recently enforced by nonlinear process for the implementation of binary optical logic [2,3]. In what follows we describe all optical polarization logic functions (XOR or XNOR) by means of noncollinear second harmonic generation (SHG) in a non phase matched Gallium Nitride (GaN) nanoslab, opening for the development of compact, integrable GaN based logic elements for parallel processing. The basic element for high speed optical signal processing is a logical gate allowing an optical signal to control a switch where a second optical signal is impinging. The basic scheme is a nonlinear interferometer in Sagnac, Mach-Zehnder, or Michelson configuration. The first such device was successfully implemented exploiting semiconductor optical amplifiers (SOA) [4,5]. Later, other nonlinear optical processes were employed and different experimental configuration were realized, based of Fiber Bragg Grating, (FBG) [6,7], or guiding structure as Kerr-effect based microring resonators [8,9]. More recently, Periodically Poled Lithium Niobate based gates were realized, relying on sum- and difference- frequency generation [10]. Other schemes are based on polarization logic, that compared with ON/OFF logic (bright or dark), require less power, thus generating less heat and being suited for cascading, as pointed out in [11], where different kinds of architectures for complete all-optical processing polarization based binary logic gates are proposed. We show here an integrable scheme of XOR/XNOR, based on non phase-matched noncollinear SHG in a medium of peculiar crystalline symmetry. This approach results useful for a parallel processing scheme. The polarization of the generated second harmonic (SH) beam is a function of the polarization of the two noncollinear pump beams. In this experimental configuration, considering for the pump beams the possibility of two polarization states, i.e. parallel ($\hat{p}$) and normal ($\hat{s}$) to the plane of incidence, there are four polarization combinations which are allowed to give SH. According to the rules shown in the following, if both pumps are $\hat{p}$ polarized or $\hat{s}$ polarized, the resulting SH is $\hat{p}$ polarized, while if the pumps beams are $\hat{s}$-$\hat{p}$ or $\hat{p}$-$\hat{s}$ the SH polarization is $\hat{s}$ polarized. The nonlinear interaction of optical signals behaves as an XOR (exclusive OR) when the optical polarizations of the two pump beams belong to a binary set as "0" ($\hat{p}$ polarization) and "1" ($\hat{s}$ polarization). When two equal bits ("0-0" or "1-1") arrive at the input of the nonlinear crystal, the output will be "0", being the optical polarization of the SH signal $\hat{p}$ polarized. On the other hand, as two different bits ("0-1" or "1-0") arrive at the input, the output bit will be "1", i.e the optical polarization of the SH signal is $\hat{s}$ polarized. By inverting the bit association to polarizations, by setting the logical "0" to $\hat{s}$ polarization and the logical "1" to $\hat{p}$ polarization, the new functionality corresponds to a XNOR (exclusive NOR) logical gate. This peculiar functionality can be obtained by using a nonlinear crystal belonging to a specific symmetry class such as 6mm (hexagonal) like Gallium nitride, for instance. The large bandgap group-III Nitride semiconductors (such as GaN, AlN and AlGaN) have had a massive impact on photonics and optoelectronics in the last decade, due to their transparency over a large range of wavelengths from the deep UV to mid/far IR [12]. As nonlinear optical material, Gallium nitride is a very promising material with large nonlinear optical coefficients comparable [13] to those of other conventional nonlinear crystals such as KDP or $LiNbO_3$. Nevertheless, the efficiency of the quadratic nonlinear interaction in bulk GaN is considered too low for practical applications, because GaN is a highly dispersive material with low birefringence, preventing the exploiting of phase-matching in bulk samples. Different GaN based structures (having thickness of several microns) [14] and quasi phase-matching in periodically poled GaN [15] were explored. Giant second harmonic signal was obtained from one-dimensional [16] and two-dimensional GaN based photonic crystals [17] where the fundamental and the generated beams are simultaneously matched with the resonant modes of the photonic crystals. In our study, we used a GaN nanoslab of only few hundred nanometres (302 nm), grown by metallo-organic chemical vapour deposition (MOCVD), with a very high quality of crystalline structure, and with a SHG configuration out of phase matching conditions. The second order susceptibility tensor of GaN presents three nonvanishing independent coefficients $\chi_{113}^{(2)}=\chi_{131}^{(2)}=\chi_{223}^{(2)}=\chi_{232}^{(2)}$, $\chi_{311}^{(2)}=\chi_{322}^{(2)}$ and $\chi_{323}^{(2)}$ [13] which are reduced to two, by taking advantage of Kleinmann symmetry rules, i.e. $\chi_{113}^{(2)}=\chi_{131}^{(2)}=\chi_{223}^{(2)}=\chi_{232}^{(2)}=\chi_{311}^{(2)}=\chi_{322}^{(2)}$ and $\chi_{323}^{(2)}$. Referring to the piezoelectric contraction [18], and being $\tilde{d}=\frac{1}{2}\tilde{\chi}^{(2)}$, the non-zero terms are thus $d_{15}=d_{24}=d_{31}=d_{32}$ and $d_{33}$.

By selecting the appropriate polarization state for both the fundamental beams and the generated one, we can differently address these components. Thus the noncollinear scheme with two input pump beams, offers very high flexibility in the handling and control of the SHG signal. Noncollinear SHG, firstly developed by Muenchausen [19] and Provencher [20] was exploited more recently by different authors. It presents some advantages, with respect to collinear SHG, as a reduced coherence length [21] and the possibility to distinguish between bulk and surface responses [22,23] thus this technique is very useful in surface and thin-film characterization [24].

## 2. Sample realization and characterization.

The epitaxial films employed in this work consist of a thin GaN layer (300 nm thick) grown at 1150°C on (0001) c-plane $Al_2O_3$ substrates in a horizontal low pressure MOCVD system (Aixtron AlX 200-RF).

Trimethyl-gallium (TMGa), trimethyl-aluminium (TMAl) and pure ammonia (NH3) were used as source materials while Palladium-diffused hydrogen was used as a carrier gas. Since sapphire and GaN are highly mismatched (around 14%), the growth of thin layer of GaN on sapphire results in low structural and morphological material quality, with a huge density of extended defects starting at the substrate interface and affecting several hundred nanometers (typically 500-600 nm) of material. This high density of defects detrimentally affects linear and nonlinear optical properties of GaN based devices [25]. In order to obtain high a better optical quality even in nm-size slab, a properly designed high temperature grown AlN nucleation layer (NL) has been studied. The 300 nm-thick GaN grown on such AlN NL was investigated in details by high resolution X-ray diffraction (HRXRD) and atomic force microscopy (AFM). As shown in Fig. 1, the full width at half maximum of the θ/2θ rocking curves for the GaN (00.2) reflection is as narrow as 164 arcsec, suggesting a low density of screw or mixed dislocations. A maximum tilt of 0.02° with respect to the c-axis (corresponding to the growth direction) has been measured. Moreover, AFM image (see the inset of Fig. 1) shows a very smooth surface with a step flow growth mode and with a roughness of only 0.3 nm measured on a scan area of 10 µm × 10 µm. A more detailed analysis on GaN grown on such a template [26] demonstrates that, since the very early stage of deposition, the AlN NL induces a fast GaN island coalescence, leading to a continuous film with a very smooth surface even for thickness as low as 50 nm. A strong reduction of the GaN epilayer mosaicity was evidenced by X-ray reciprocal space mapping, with the disorder induced by the large mismatch being concentrated only in the first 50 nm.

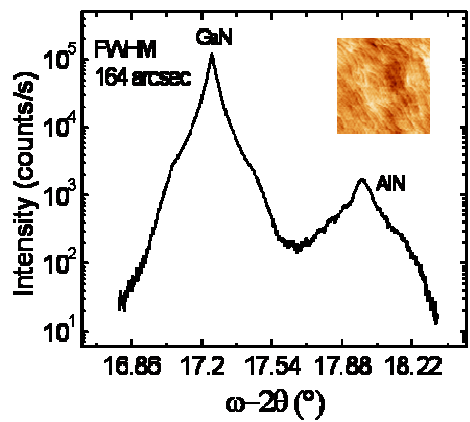

**Fig. 1. X-Ray spectra showing the GaN peak at 17.25 ° with a FWHM of 164 arcsec and the AlN peak at 18 °. Inset: AFM image of GaN slab surface on a 5x5 μm area: rms is less than 0.3 nm.**

## 3. Experimental configuration.

We carried out the SHG measurements by means of a noncollinear rotational Maker fringes technique working in the transmission scheme, as reported in Fig. 2. In our experiments the output of a mode-locked Ti:Sapphire laser system tuned at λ=830 nm (76 MHz repetition rate, 130 fs pulse width and average power of 500 mW ) was split into two beams of the same intensity using a 50/50 beam splitter. The polarization of both beams can be controlled by two identical rotating half wave plates, which were carefully checked not to give nonlinear contribution since the two collimating lenses, 150 mm focal length, are placed thereafter. The tightly focused beams were sent to intersect in the focus region with an aperture angle of 18° with respect to each other. The sample was placed onto a motorized combined translation and rotation stage which allowed the variation of the sample rotation angle, $\alpha$, with a resolution of 0.05 degrees as well as the z-scan of the sample position within the two beam overlap. Meanwhile, the temporal overlap of the incident pulses was automatically controlled by an external delay line.

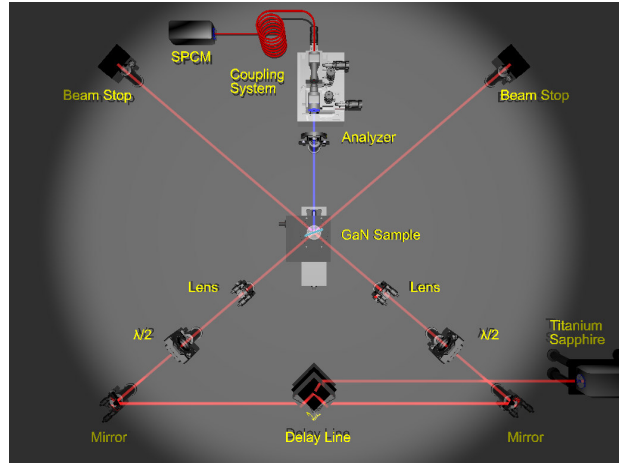

**Fig. 2. Sketch of the non-collinear SHG experimental set-up .**

Considering the two beams, tuned at $\omega_1=\omega_2$, impinging on the sample at two different incidence angles, namely $\alpha_1$ and $\alpha_2$, measured with respect to $\alpha=0$, the resulting nonlinear polarization is composed by three waves oscillating at $2\omega_1$, $2\omega_2$, and $\omega_1+\omega_2$, respectively. While the first two waves are nearly collinear with the incident beams, the third one has a wave vector directed almost along the bisector among the two input beams, i.e. forming an internal angle with respect to the normal, given by the conservation of momentum tangential component [24]. This latter beam was collected with an objective and focused on to a monomodal optical fiber coupled with a photon counting detector (SPCM). Here, a set of optical low pass filters further suppress any residual and scattered light at fundamental frequency and an analyzer allows to select the desired SH polarization state. Finally, all parts of the experiment were automatically controlled.

As mentioned in the introduction, for pump beams having same polarization, either $\hat{p}$-$\hat{p}$ or $\hat{s}$-$\hat{s}$, the generated beam is $\hat{p}$-polarized, as well as in the collinear case. In case of crossed polarizations, $\hat{p}$-$\hat{s}$ or $\hat{s}$-$\hat{p}$, the resulting SH is $\hat{s}$ polarized. This process is exhaustively illustrated by the analytical expression of the effective susceptibility, $d_{\text{eff}}(\alpha)$, in the four different combinations once the $\tilde{d}$ tensor corresponding to GaN is assumed:

$$\begin{aligned}
d_{\text{eff}}^{\text{ppP}} &= -\cos(\alpha_{2\omega})d_{24}\left[\cos(\alpha_1')\sin(\alpha_2') + \cos(\alpha_2')\sin(\alpha_1')\right] + \\
&\quad + \sin(\alpha_{2\omega})\left[-d_{32}\cos(\alpha_1')\cos(\alpha_2') - d_{33}\sin(\alpha_2')\sin(\alpha_1')\right] \\
d_{\text{eff}}^{\text{ssP}} &= -\sin(\alpha_{2\omega})d_{31} \\
d_{\text{eff}}^{\text{psS}} &= -d_{15}\sin(\alpha_1') \\
d_{\text{eff}}^{\text{spS}} &= -d_{15}\sin(\alpha_2')
\end{aligned} \quad (1)$$

where $\alpha'_1, \alpha'_2$ are the internal angles of the two impinging pump beams inside the sample, given by Snell's law and $\alpha_{2\omega}$ is the angle of emission of the noncollinear SH inside the crystal. Experimental results of the generated signal for the four combinations of polarization states given in Eqs. (1) are shown in Fig.3, where the SH signal is measured as a function of the sample rotation angle and position within the z-scan (only positive angles are reported). The different shape of the signal contour plots reflects the different behavior of GaN nonlinear optical tensor under different condition of excitation.

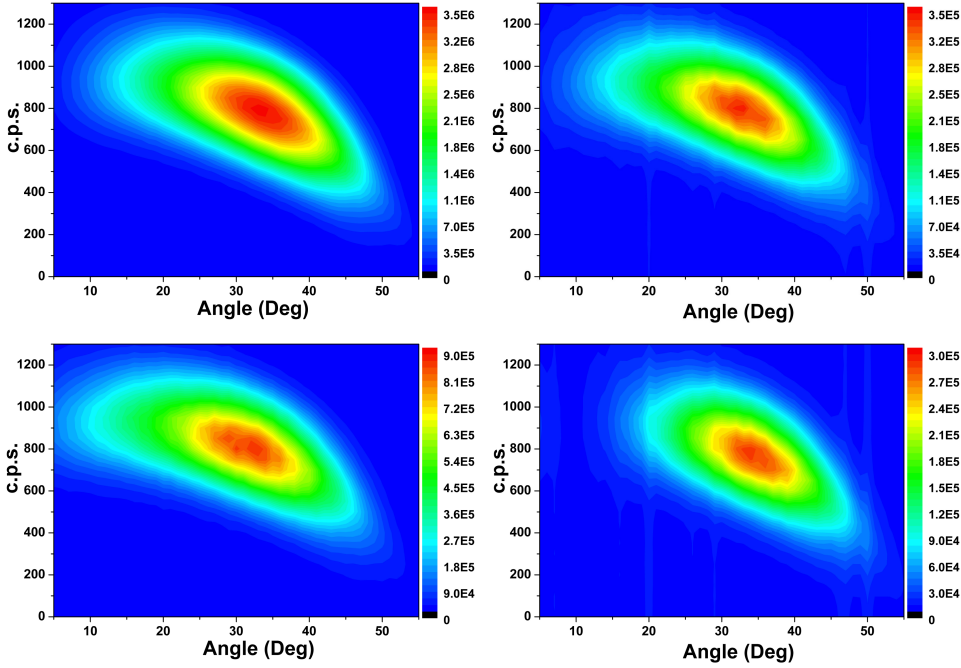

**Fig. 3. Contour plot of the generated second harmonic signal as a function of the incidence angle and sample position within the z-scan: (a) the two pump beams have same polarization state ($\hat{p}$-$\hat{p}$); (b) the two pump beams have same polarization state ($\hat{s}$-$\hat{s}$); (c) crossed polarization state ($\hat{s}$-$\hat{p}$); (d) crossed polarization state ($\hat{p}$-$\hat{s}$).**

Excepting for the absolute value (each plot represents the number of photon counting per second with different level of the intensity), the first two plots (Fig. 3a, and Fig. 3b), corresponding to the $\hat{p}$ polarized SH signal measured when the pumps are $\hat{p}$-$\hat{p}$ and $\hat{s}$-$\hat{s}$ polarized, respectively, display a similar shape for the SH trace over incidence angle and z-scan. Unlikely, when the polarization states of pumps are crossed, resulting in $\hat{s}$ polarized SH signal, there are evident differences in the contour plot shape (Fig. 3c and Fig. 3d). The position of the two beams, differently polarized, with respect to the sample surface is no more symmetrical, according to the representation of the nonlinear optical coefficient given in the Eq.(1). For $\hat{p}$-$\hat{s}$ excitation the signal versus rotation angle always retains a nonzero value, when the sample rotation angle approaches the zero value, while in the reverse situation, i.e. $\hat{s}$-$\hat{p}$ excitation, it quickly reaches the zero value at about 9° of rotation angle, where one of the two beams is normally incident onto sample surface. The peculiarity of the angular behavior of the nonlinear optical response can be further evidenced by analyzing the SH curve as a function of positive sample rotation angles. Angular scan were obtained by rotating the sample in the range -55°-55°, excluding the angles surrounding the 0° (±5°) in order to avoid back reflections into the laser system. The outcoming curves are plotted in Fig. 4. Red and black curves always represent $\hat{p}$ and $\hat{s}$ polarized SHG signals, respectively. In Fig. 4a both pumps have the same $\hat{p}$ polarization, while in Fig. 4b they're both $\hat{s}$ polarized. It can be seen that there is no significant $\hat{s}$ polarized SHG signal is measured when the pump beams have the same polarization. The form of the effective nonlinear optical susceptibility, results in two balanced curves for the case of similar polarization $\hat{p}$-$\hat{p}$ and $\hat{s}$-$\hat{s}$, as well as a signal asymmetry for the $\hat{s}$-$\hat{p}$ and $\hat{p}$-$\hat{s}$ excitation. Given two pump beams with crossed polarizations ($\hat{s}$-$\hat{p}$ and $\hat{p}$-$\hat{s}$ excitation) the SH signal is mainly $\hat{s}$-polarized, while a very low $\hat{p}$-polarized SH signal is detectable (Fig. 4c and Fig. 4d). In all cases we recognize that the maximum efficiency of the process occurs at 34°, and if we now attribute to each polarization state the bit "0" or "1" we can easily find that all results reveal the logic functionality described above as XOR (or) XNOR with the optimum "working point" located at 34°.

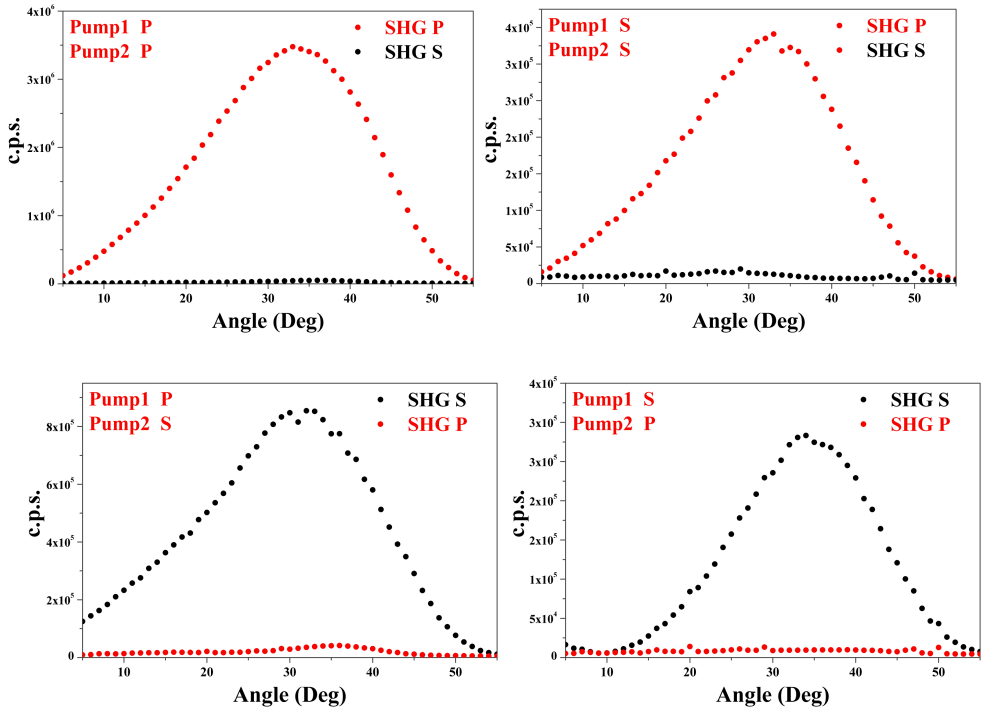

**Fig. 4. Generated signal as a function of the incidence angle. The analyzer that detect the outcoming second harmonic signal is set to $\hat{p}$ (red curve) and $\hat{s}$ (black curve) polarization, respectively: (2a) The two pump beams have same polarization state ($\hat{p}$-$\hat{p}$); (2b) Pumps with ($\hat{s}$-$\hat{s}$) input polarization; (2c) The two pump beams have crossed polarization state; Pumps with ($\hat{s}$-$\hat{p}$) polarization respectively; (2d) Pumps with ($\hat{p}$-$\hat{s}$) polarization respectively.**

In spite of the non phase-matching conditions, the GaN wurtzite crystal structure [12, 13] leads to SHG, with very low input pump power, in all the possible polarization combinations which are allowed by its nonlinear optical tensor in the SHG noncollinear configuration. As a consequence, we can handle the output polarization states of the generated signal by choosing an appropriate polarization combination for the two pump beams.

## 4. Conclusions

We have experimentally demonstrated an optical logical functionality as XOR or XNOR based on polarization states from a very thin GaN nanoslab by means of a non phase-matched noncollinear second harmonic generation. The logical bit is associated to the polarization state, that can be handled thank to the peculiar crystaline symmetry of GaN. The scheme, which is based on a nonlinear process, should result useful for parallel logic. The proposed scheme is characterized by high simplicity of realization and can be easily integrated giving the possibility of parallel optical functionalities with GaN.